\documentstyle[11pt,epsfig,menu99]{article}

\def\lapx{\,\,\lower 2pt \hbox{$\buildrel<\over{\scriptstyle{\sim}}$}\,\,}
\begin{document}
    \setlength{\baselineskip}{2.6ex}
\title{Isospin Violation and the Proton's Strange Form Factors}
\author{Randy Lewis and Nader Mobed \\
{\em Department of Physics, University of Regina, Regina, SK, Canada S4S 0A2}}

\maketitle

\begin{abstract}
\setlength{\baselineskip}{2.6ex}
The strange form factors of the proton are basic to an understanding of
proton structure, and are presently the focus of many experiments.
Before the strangeness effects can be extracted from data,
it is necessary to calculate and remove effects due to isospin violation,
which exist independently of the strange quark but which contribute
nevertheless to the experimentally measured ``strange'' form factors.
A discussion of the isospin violating contributions to vector form factors
is given here in the context of heavy baryon chiral perturbation theory.
\end{abstract}

\setlength{\baselineskip}{2.6ex}

\section*{INTRODUCTION}

The interaction between a proton and a neutral weak boson ($Z^0$) involves
form factors which are related to the familiar electromagnetic form factors via
the standard electroweak model.
For example, the proton's neutral weak vector form factors are
\begin{equation}\label{main}
   G_X^{p,Z}(q^2) =
   \frac{1}{4}[G_X^p(q^2)-G_X^n(q^2)]-G_X^p(q^2){\rm sin}^2\theta_W
   -\frac{1}{4}G_X^s(q^2),~~~X=E,M~,
\end{equation}
where $G_{E,M}^p(q^2)$ and $G_{E,M}^n(q^2)$ are the usual electromagnetic form 
factors of the proton and neutron respectively, and $G_{E,M}^s(q^2)$ are 
called the proton's strange electric and magnetic form factors.
Using Eq. (\ref{main}), an experimental measurement of $G_X^{p,Z}(q^2)$ 
leads to a determination of $G_X^s(q^2)$, which provides
information about the effects of strange quarks in the proton.

The first measurement of $G_M^{p,Z}(q^2)$ was reported two years ago by the 
SAMPLE Collaboration\cite{SAMPLE}, and led to
\begin{equation}\label{SAMeq}
G_M^s(0.1{\rm GeV}^2) = 0.23 \pm 0.37 \pm 0.15 \pm 0.19~.
\end{equation}
A linear combination of strange electric and magnetic form factors has been
measured by the HAPPEX Collaboration\cite{HAPPEX}:
\begin{equation}\label{HAPeq}
[G_E^s+0.39G_M^s](0.48{\rm GeV}^2) = 0.023 \pm 0.034 \pm 0.022 \pm 0.026~.
\end{equation}
Further efforts are underway by various groups.\footnote{See 
in particular the second SAMPLE measurement, Ref. \cite{SAMPLE2}, which
appeared after the MENU99 conference.  Using a calculation of electroweak
corrections as input, they find 
$G_M^s(0.1{\rm GeV}^2) = +0.61 \pm 0.17 \pm 0.21$.}

It is important to recall that $G_E^s(q^2)$ and $G_M^s(q^2)$ contain more 
than just 
strangeness effects.  Even in a world of only two flavours (up and down) 
$G_{E,M}^s(q^2)$ would be
nonzero due to isospin violation.  Thus, the true effects of strange quarks
can only be extracted from an experimental determination of $G_{E,M}^s(q^2)$
if isospin violating effects can be calculated.

Dmitra\v{s}inovi\'{c} and Pollock\cite{DmiPol}, and also Miller\cite{Mil},
have studied the isospin violating contribution to $G_M^s(q^2)$
within the nonrelativistic constituent quark model.
Ma has used a light-cone meson-baryon fluctuation model.\cite{Ma}
More recently, a model-independent study of isospin violating effects
(using heavy baryon chiral perturbation theory) has been published\cite{LewMob}
and it is this work which will be emphasized below, after a brief review
of attempts to calculate the authentic strange quark effects.

\section*{THE STRANGENESS CONTRIBUTIONS TO \boldmath{$G_{E,M}^s(q^2)$}}

Many attempts have been made to calculate the contribution of strange quarks
to the ``strange'' electric and magnetic form factors, $G_{E,M}^s(q^2)$.
In principle a lattice QCD calculation could give the definitive answer,
and an exploratory calculation has been performed in the quenched 
theory.\cite{latQCD}
The errors due to finite lattice spacing, finite lattice volume and 
quenching are not yet known, but the existing results,
$\left<r_s^2\right>_E \equiv 6{\rm d}G_E^s(0)/{\rm d}q^2=
-0.06\rightarrow-0.16{\rm fm}^2$ and $G_M^s(0) = -0.36\pm0.20$,
still provide important inputs to the discussion.

One might consider using chiral perturbation theory to calculate the 
strangeness contributions to $G_{E,M}^s(q^2)$, but both form factors have
a free parameter 
at their first nonzero order in the chiral expansion,
so the magnitude of neither form factor can be predicted from chiral 
symmetry alone.  However, two experimental inputs are sufficient to fix
these parameters, and chiral symmetry does determine
the $q^2$-dependence of the form factors at leading chiral order.  This 
tact has been taken by the authors of Ref. \cite{strangeChPT}, 
who use the SAMPLE and HAPPEX measurements as input.

Beyond lattice QCD and chiral perturbation theory, there are many
models and dispersion relation methods which have been employed in the
effort to determine the strange quark contributions to $G_{E,M}^s(q^2)$.
(The authors of Refs. \cite{latQCD,strangeChPT,models} have collected some 
predictions from the literature.)
The various methods lead to differing results.  For $G_M^s(0)$, most
predictions lie in the range
\begin{equation}
   -0.5 \lapx G_M^s(0) \lapx +0.05~,
\end{equation}
and it has often been noted that this tendency toward a negative number does
not seem to be supported by the experimental data, 
Refs. \cite{SAMPLE,HAPPEX,SAMPLE2}.
Predictions for the magnitude and sign of $\left<r_s^2\right>_E$ also span 
a large range.

A precise experimental measurement would help to distinguish between the 
various models of strangeness physics, but only after the isospin violating 
contribution has been calculated and subtracted.

\section*{THE ISOSPIN VIOLATING CONTRIBUTIONS TO \boldmath{$G_{E,M}^s(q^2)$}}

In a world with no strange quark, $G_E^s(q^2)$ and $G_M^s(q^2)$ do not vanish.
Instead,
\begin{equation}
G_X^s(q^2) \rightarrow G_X^{u,d}(q^2) 
{\rm ~as~the~strange~quark~decouples,~} (X=E,M)
\end{equation}
where $G_E^{u,d}(q^2)$ and $G_M^{u,d}(q^2)$ are isospin violating quantities.
If both the strange and isospin violating components of $G_{E,M}^s(q^2)$ are 
small, then 
contributions which are both isospin violating {\it and\/} strange are
doubly suppressed.
The following discussion considers
$G_{E,M}^{u,d}(q^2)$ in a strange-free world.

Constituent quark model calculations have led to a vanishing result 
for $G_M^{u,d}(0)$ and a very mild $q^2$ dependence:
$-0.001<G_M^{u,d}(-0.25{\rm GeV}^2)<0$\cite{DmiPol,Mil}.
There is no symmetry which would force $G_M^{u,d}(0)$ to vanish exactly, but
perhaps the constituent quark model is trying to anticipate a ``small'' result.
A light-cone meson-baryon fluctuation model permits a large range,
$G_M^{u,d}(0) = 0.006 \rightarrow 0.088$.\cite{Ma}

Heavy baryon chiral perturbation theory (HBChPT) is a natural tool for 
the study of $G_{E,M}^{u,d}(q^2)$.  It is a model-independent approach which
employs a systematic expansion in small momenta ($q$), small pion masses 
($m_\pi$), small QED coupling ($e$), large chiral scale ($4\pi{F_\pi}$) 
and large nucleon masses ($m_N$).  It is appropriate to use 
$O(q) \sim O(m_\pi) \sim O(e)$ with $4\pi{F_\pi} \sim m_N$, and then the HBChPT 
Lagrangian can be ordered as a single expansion,
\begin{equation}\label{LpiN}
   {\cal L}_{\rm HBChPT} = {\cal L}^{(1)} + {\cal L}^{(2)} + {\cal L}^{(3)} 
                         + {\cal L}^{(4)} + {\cal L}^{(5)} + \ldots~.
\end{equation}
For the explicit form of the Lagrangian, see Ref. \cite{LewMob} and references
therein.  For the present discussion, it is simply noted that
${\cal L}^{(1)}$ contains parameters $g_A$, $F_\pi$ and $e$;
${\cal L}^{(2)}$ contains 11 parameters (7 strong and 4 
electromagnetic); ${\cal L}^{(3)}$ contains 43 parameters;
${\cal L}^{(4)}$ contains hundreds of parameters and
${\cal L}^{(5)}$ has even more.
Happily, it will be shown that $G_E^{u,d}(q^2)$ is parameter-free at its
first nonzero order, and $G_M^{u,d}(q^2)$ is parameter-free at its first
and second nonzero orders except for a single additive constant.

The coupling of a vector current (e.g. $Z^0$) to a nucleon begins at first
order in HBChPT, ${\cal L}^{(1)}$, but is isospin conserving.
To be precise, recall the usual notation,
\begin{equation}\label{defFs}
\left<N(\vec{p}+\vec{q})|\bar{f}\gamma_\mu{f}|N(\vec{p})\right> \equiv
    \bar{u}(\vec{p}+\vec{q})\left[\gamma_\mu{F}_1^f(q^2)
    +\frac{i\sigma_{\mu\nu}q^\nu}{2m_N}{F}_2^f(q^2)\right]u(\vec{p})~,
\end{equation}
where $f$ denotes a particular flavour of quark.
The Sachs form factors for that flavour are
\begin{equation}
G_E^f(q^2) = F_1^f(q^2)+\frac{q^2}{4m_N^2}F_2^f(q^2)~,~~~~~
G_M^f(q^2) = F_1^f(q^2)+F_2^f(q^2)~.
\end{equation}

An explicit calculation using 
${\cal L}^{(1)}+{\cal L}^{(2)}+{\cal L}^{(3)}$
leads to isospin violating vector form factors which vanish exactly.
At first glance this might seem surprising, but it can be readily 
understood as follows.
An isospin violating factor, 
such as $(m_n-m_p)/m_p$, is suppressed by two HBChPT orders.
Moreover, the $F_2$ term in Eq. (\ref{defFs}) has an extra explicit
$1/m_N$ suppression factor, so isospin violating $F_2$ terms cannot
appear before ${\cal L}^{(4)}$.  Meanwhile, $F_1$ is constrained
by Noether's theorem (QCD's flavour symmetries: upness and downness)
to be unity plus momentum-dependent corrections, and dimensional analysis
therefore requires a large scale, $m_N$ or $4\pi{F_\pi}$, in the denominator
of all corrections.
This demonstrates that both $G_E^{u,d}(q^2)$ and $G_M^{u,d}(q^2)$ vanish
in HBChPT until the fourth order Lagrangian: ${\cal L}^{(4)}$.

A leading order (LO) calculation of $G_E^{u,d}(q^2)$ or $G_M^{u,d}(q^2)$ 
involves tree-level terms from ${\cal L}^{(4)}$ plus one-loop 
diagrams built from ${\cal L}^{(1)}+{\cal L}^{(2)}$.
Referring to Ref. \cite{LewMob} for details of the calculation and
renormalization, the results are
\begin{eqnarray}
\left.G_E^{u,d}(q^2)\right|_{\rm LO} &=& 
-\frac{4\pi{g}_A^2m_{\pi^+}}{(4\pi{F})^2}
         (m_n-m_p)\left[1-\int_0^1{\rm d}x\,\frac{1-(1-4x^2)q^2/m_{\pi^+}^2}
         {\sqrt{1-x(1-x)q^2/m_{\pi^+}^2}}\right], \\
\left.G_M^{u,d}(q^2)\right|_{\rm LO} &=& {\rm constant} 
- \frac{16g_A^2m_N}{(4\pi{F})^2}(m_n-m_p)
               \int_0^1{\rm d}x\,{\rm ln}\left(1-x(1-x)\frac{q^2}{m_{\pi^+}^2}
               \right).
\end{eqnarray}
Notice that the electric form factor contains no unknown parameters, and the
magnetic form factor has only a single parameter (an additive constant).
The LO results for $G_E^{u,d}(q^2)$ and $G_M^{u,d}(q^2)-G_M^{u,d}(0)$ are 
plotted in Fig. 1.  
The contribution of isospin violation to $\left<r_s^2\right>_E$ is 
$6{\rm d}G_E^{u,d}(0)/{\rm d}q^2\approx+0.013{\rm fm}^2$.

Consider next-to-leading 
order (NLO).  Here, one expects tree-level terms from ${\cal L}^{(5)}$ 
plus one- and two-loop diagrams built from lower orders in the Lagrangian.
Since small HBChPT expansion parameters without uncontracted Lorentz indices
come in pairs (e.g. $q^2$, $m_\pi^2$, $e^2$), the ${\cal L}^{(5)}$ counterterms
can contribute to $F_1$ but not to $F_2$.  Thus $G_M^{u,d}(q^2)$ is independent
of these parameters at NLO, although $G_E^{u,d}(q^2)$ is not.

It is also found that no two-loop diagrams contribute to $G_M^{u,d}(q^2)$ at 
NLO, although in principle they could have.  Furthermore, unknown coefficients
from ${\cal L}^{(3)}$ are also permitted to appear within loops,
but none of them actually contribute.
This means that the NLO corrections to $G_M^{u,d}(q^2)$ are basic one-loop
diagrams.  The explicit result is given in Ref. \cite{LewMob}.  It needs
to be stressed that the NLO contribution is parameter-free; the only new
quantities (with respect to LO) are the well-known nucleon magnetic moments.

The LO+NLO result for $G_M^{u,d}(q^2)-G_M^{u,d}(0)$ is shown in Fig. 2.
Notice that the NLO corrections serve to soften the $q^2$-dependence.
The NLO correction to $G_M^{u,d}(0)$ is
\begin{equation}\label{NLO}
\left.G_M^{u,d}(0)\right|_{LO+NLO} - \left.G_M^{u,d}(0)\right|_{LO} = 
\frac{24\pi{g}_A^2m_{\pi^+}}{(4\pi{F})^2}(m_n-m_p)\left(\frac{5}{3}-\mu_p
-\mu_n\right) \approx 0.013~.
\end{equation}

The value of $G_M^{u,d}(0)$ itself is not determined by chiral symmetry alone,
and it receives contributions from physics other than the ``pion cloud''
of HBChPT (consider, for example, isospin violation due to vector mesons).
The pion cloud contribution to $G_M^{u,d}(0)$ is estimated in 
Ref. \cite{LewMob} via a physically-motivated cutoff in HBChPT, and is
comparable in size to the NLO contribution of Eq. (\ref{NLO}).

The full result for the pion cloud contribution to $G_M^{u,d}(q^2)$ is 
shown in Fig. 2 with error bands to reflect truncation of the HBChPT expansion:
the narrow band assumes $|{\rm NNLO}|\sim|{\rm NLO}|\cdot{m}_\pi/m_N$ 
and the wide band assumes $|{\rm NNLO}|\sim|{\rm NLO}|/2$.

\parbox{8cm}{
   \epsfig{figure=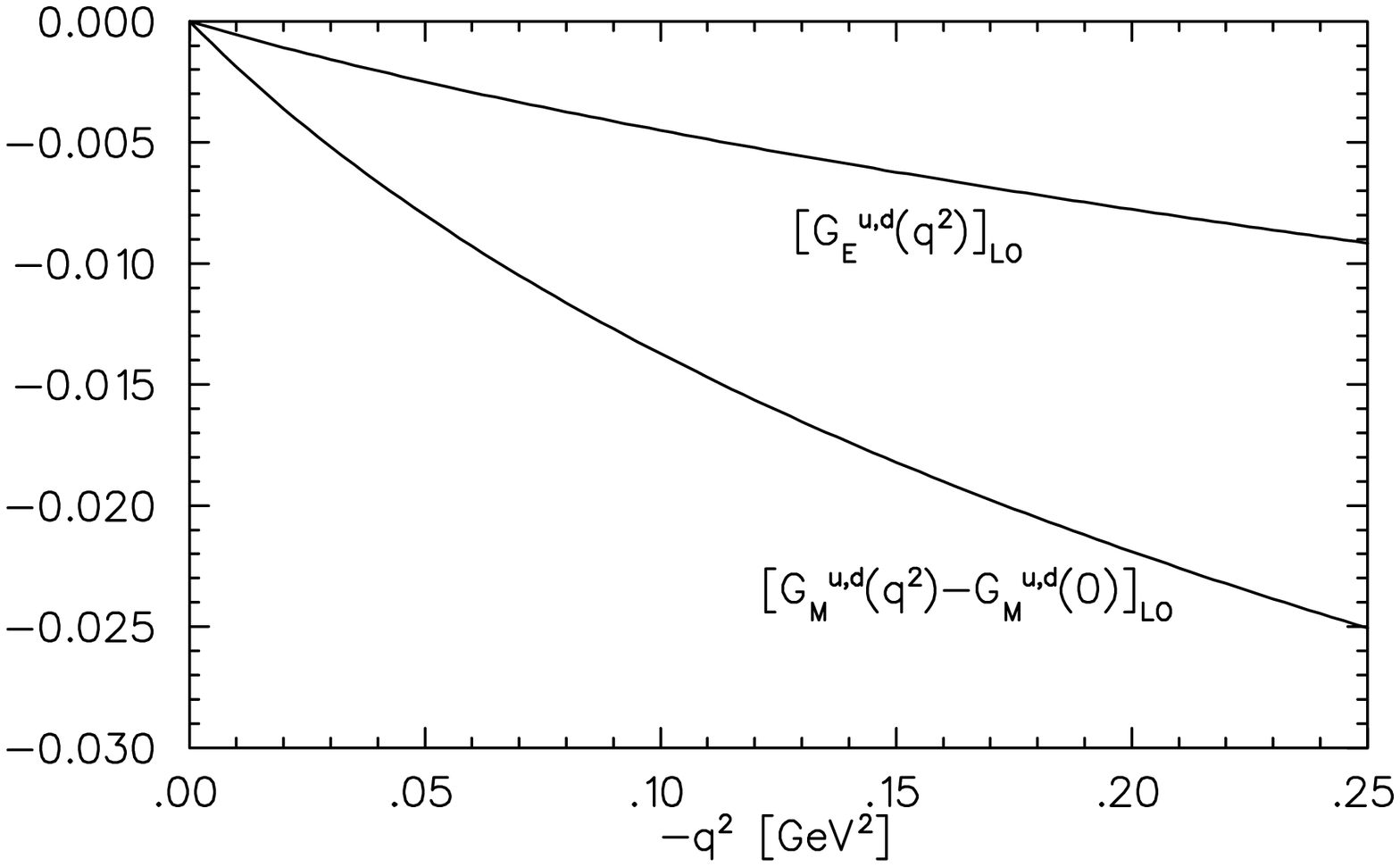,width=7cm,height=6cm}
   \noindent
   \parbox{7cm}
   {\small \setlength{\baselineskip}{2.6ex} Fig.~1. Parameter-free results
   for $G_E^{u,d}(q^2)$ and
   $G_M^{u,d}(q^2)-G_M^{u,d}(0)$ at LO in HBChPT.}}
\parbox{8cm}{\vspace*{3mm}
   \epsfig{figure=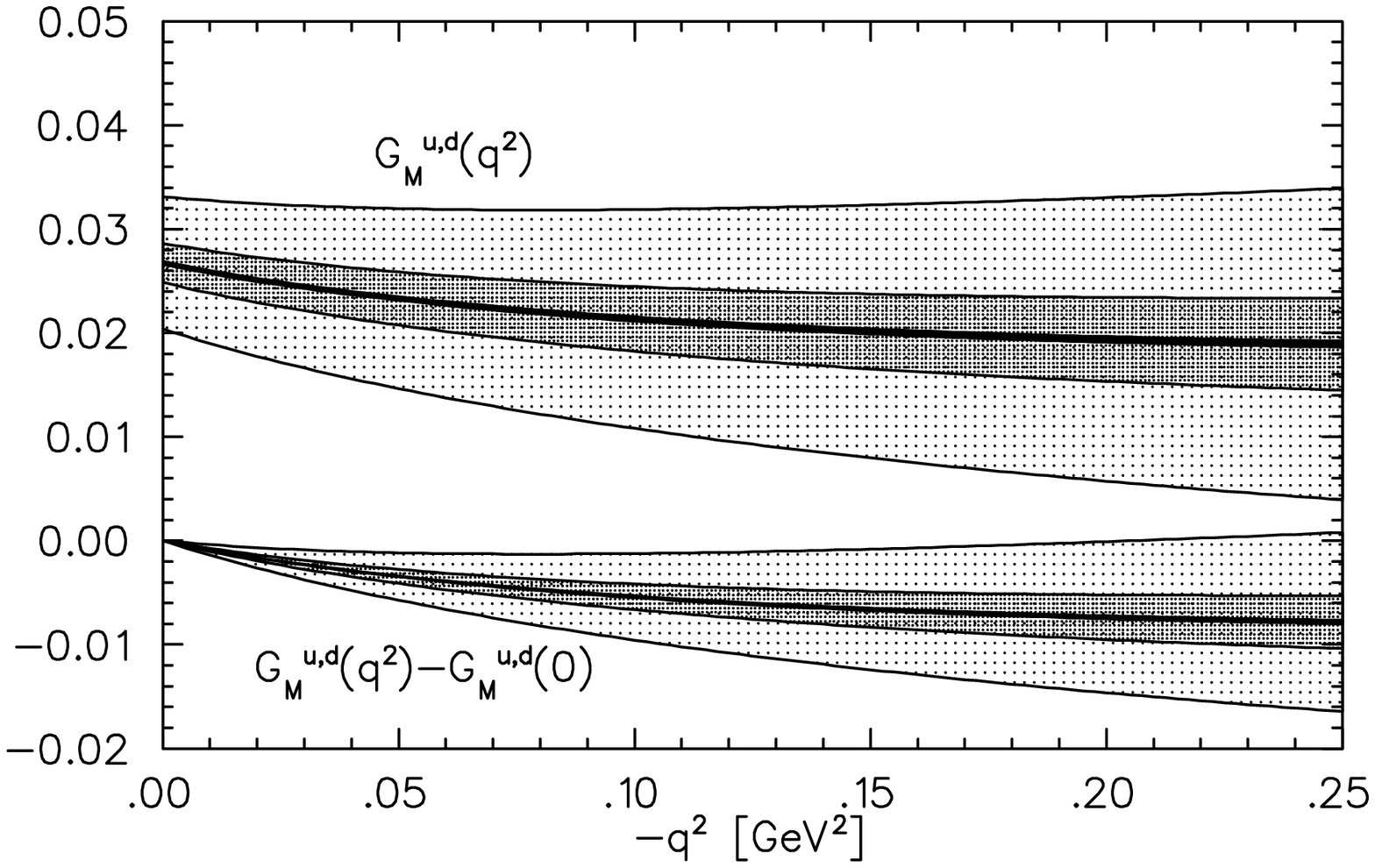,width=7cm,height=6cm}
   \noindent
   \parbox{7cm}
   {\small \setlength{\baselineskip}{2.6ex} Fig.~2. The pion cloud contribution
   to $G_M^{u,d}(q^2)$ at LO+NLO, with uncertainties due to truncation of
   the HBChPT expansion.}}

\section*{CONCLUSIONS}

The strange vector form factors of the proton are basic to an understanding
of proton structure.  The contribution due to strange quarks has proven to
be a theoretical challenge.  Isospin violation also contributes to the
so-called ``strangeness'' form factors, and this contribution must be
calculated and subtracted from experimental data before the strange quark
contribution can be identified.

The present work indicates that chiral symmetry is of great value for
discussions of the isospin violating effects.  Despite the large number
of parameters in the Lagrangian, $G_E^{u,d}(q^2)$ is parameter-free at
leading order, and $G_M^{u,d}(q^2)$ has only one ($q^2$-independent)
parameter at leading order, and no parameters at next-to-leading order.

The isospin violating effects computed here are large compared to some models
of the strange quark effects, but small compared to other models.
The experimental results for the full
``strangeness'' form factors in Eqs. (\ref{SAMeq}) and (\ref{HAPeq})
are not precise enough to indicate their size relative to the isospin
violating contributions found in this work.
It will be interesting to see what future experiments reveal.

\vspace{5mm}

This work was supported in part by the Natural Sciences and Engineering
Research Council of Canada.

\bibliographystyle{unsrt}

\end{document}